\documentclass[%
amsmath,
amssymb,
]{revtex4-1}

\usepackage{graphicx}
\usepackage{dcolumn}
\usepackage{bm}

\usepackage{amsmath}
\usepackage{color}
\usepackage[normalem]{ulem}
\usepackage{balance}
\usepackage[T1]{fontenc}
\usepackage{caption}

\makeatletter
\renewcommand*{\@fnsymbol}[1]{\ifcase#1\or*\else\@arabic{#1}\fi}
\makeatother

\begin{document}
\title{Classic Harmonic Oscillator Model of Coupled Metal Nanoparticles with Arbitrary Configuration in Plane}

\author{Yuqing Cheng}\thanks{Corresponding author: yuqingcheng@ustb.edu.cn}
\affiliation{School of Mathematics and Physics, University of Science and Technology Beijing, Beijing 100083, China}%

\begin{abstract}
\textbf{ABSTRACT: }A classic harmonic oscillator model is developed to investigate the optical properties of coupled metal nanoparticles (MNPs) with arbitrary configuration in plane. The coupling coefficients are derived from classical electrodynamics. Using this model, we can easily obtain the spectra of coupled MNPs varying with the configurations of the system and the polarizations of external light. Furthermore, the far field electric field distributions of different configurations are revealed. This model is an extension of our previous works which only discuss the parallel and vertical excitations for dimer. It is useful to related applications
\end{abstract}
\maketitle

\section{\label{sec:Introduction}Introduction}
The optical properties of coupled metal nanostructures are worth investigating due to their excellent behavior in controlling light at the subwavelength scale. Therefore, they have attracted the attention of researchers due to their potential in the applications.
Models have been developed to explain the coupling effect.\cite{couple2,couple3,couple4,couple5,couple6}
In our previous work, the coupling models of MNPs with dimer configuration are developed.\cite{couple0,couple1} Two or three coupling coefficients are obtained under the first order approximation, and the photoluminescence and scattering spectra of the dimer are investigated in detail. However, theses models can only solve the parallel and vertical excitation problems, and cannot be used to explain arbitrary configuration, especially for many--body system.

In this study, we present a classic oscillator model to reveal the optical properties of arbitrary configurations. Start from the two--body system, the coupling coefficients are derived; then extend them to the many--body system in plane to reveal the behaviors of the arbitrary complicated configuration.

\section{\label{sec:Model}Model}
We extend the previous coupling models which can only solve the simple configurations of parallel and vertical excited dimer to a many-body coupling case with arbitrary configuration, the optical properties of which can be evaluated by this extended model. In this model, we treat each MNP as an individual oscillator which oscillates along both $x$ and $y$ directions simultaneously. All the MNPs are in the same $xy$ plane. As a many-body problem, the coupling among the MNPs is much more complicated than the parallel or vertical excited dimer case.

\begin{figure}[b]
\includegraphics[width=0.48\textwidth]{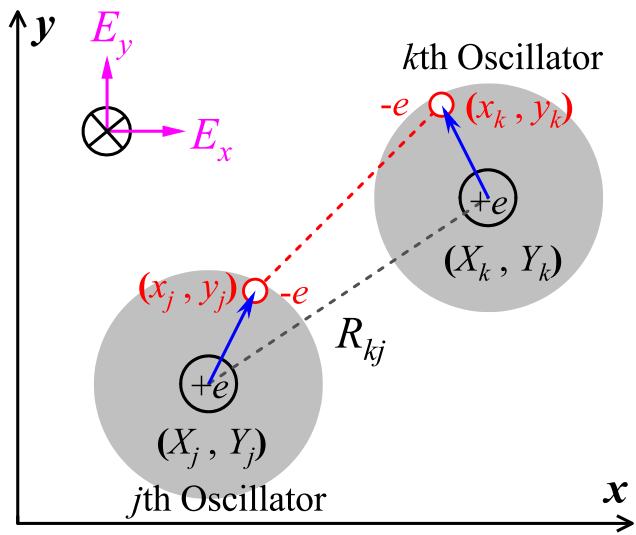}
\caption{\label{fig:scheme} Schematic of the model, taking two oscillators (big circles filled with gray), the $j$th and the $k$th, as an example. The ions with positive charges ($+e$) are steady at the positions $(X_j,Y_j)$ and $(X_k,Y_k)$. The electrons with negative charges ($-e$) oscillate with the positions $(x_j,y_j)$ and $(x_k,y_k)$ relative to the positions of their own ions. $R_{kj}$ is the distance between the two ions.
}
\end{figure}

Fig. \ref{fig:scheme} shows the schematic of the model, illustrating two oscillators as an example. The oscillator is made of ion with positive charge and electron with negative charge. The ions are assumed to be steady, and the absolute coordinates of the $j$th ion and $k$th ion are $\textbf{R}_j=(X_j,Y_j)$ and $\textbf{R}_k=(X_k,Y_k)$, respectively, thus the distance between them $R_{kj}=\left|\textbf{R}_k-\textbf{R}_j \right|$; while the electrons oscillates around their own ions, and $(x_j,y_j)$ and $(x_k,y_k)$ represent the coordinates at which the electrons deviate from their equilibrium positions, respectively. Therefore, we can define $\dot{x}_j$ and $\ddot{x}_j$ as the $x$ component of the velocity and the acceleration of the $j$th electron, and define $\dot{y}_j$ and $\ddot{y}_j$ as the $y$ component ones. The external electric field with (circular) frequency $\omega_{ex}$ propagates along $z$ direction (perpendicular to the page) with its $x$ and $y$ polarization components $E_x$ and $E_y$, respectively.

For a many--body system with $n$ MNPs, the equations satisfied by the electron are written as:
\begin{equation}
\centering
\begin{aligned}
&\ddot{x}_j+\beta_{0xj}\dot{x}_j+\omega_{0xj}^2 x_j
+\frac{e}{m_e}\sum_{k\neq j}^{n}E_{kjx}
=C_{jx}\mathrm{exp}(-\mathrm{i}\omega_{ex}t),\\
&\ddot{y}_j+\beta_{0yj}\dot{y}_j+\omega_{0yj}^2 y_1
+\frac{e}{m_e}\sum_{k\neq j}^{n}E_{kjy}
=C_{jy}\mathrm{exp}(-\mathrm{i}\omega_{ex}t),\\
&j=1,~2,~...,~n.
\end{aligned}
\label{eq:xy0}
\end{equation}
Here, $C_{jx}=-e E_{jx}/m_e, C_{jy}=-e E_{jy}/m_e$, $e$ is the elemental charge, and $m_e$ is the electron mass; $E_{jx}$ and $E_{jy}$ refer to the external electric field of $x$ and $y$ polarization felt by the $j$th oscillator, respectively; $E_{kjx}$ and $E_{kjy}$ refer to the electric field of $x$ and $y$ polarization felt by the $j$th oscillator introduced by the $k$th oscillator.
In order to calculate $E_{kjx}$ and $E_{kjy}$, it is necessary to take into account the formula that express the electric field introduced by the moving charge:\cite{Griffiths}
\begin{equation}
\begin{aligned}
&\textbf{E}=\frac{q}{4\pi \varepsilon_0}\frac{r}{\left(\textbf{r}\cdot \textbf{u}\right)^3}
\left[\left(c^2-v^2\right)\textbf{u}+\textbf{r}\times \left(\textbf{u}\times \textbf{a}\right)
\right],\\
&\textbf{B}=\frac{1}{cr}\textbf{r}\times\textbf{E}.
\end{aligned}
\label{eq:Ert}
\end{equation}
Here, $\varepsilon_0$ is the permittivity of free space, and $c$ is the speed of light in vacuum; $q$, $v=\left|\textbf{v}\right|$, and \textbf{a} are the charge, velocity, and acceleration of a moving point charge, and \textbf{r} is the displacement vector from the point charge to field point; $\textbf{u}\equiv c\textbf{r}/r-\textbf{v}$.
Back to our problem, notice that $\textbf{E}_{kj}$ is the electric field introduced by both the moving electron and the steady ion of the $k$th oscillator, therefore we can obtain its $x$ and $y$ components derived from Eq. \ref{eq:Ert}:
\begin{equation}
\centering
\begin{aligned}
&E_{kjx}\cong \frac{-N_k e}{4\pi\varepsilon_0 R_{kj}^2}
\left[\left( \frac{3\left(\Delta X_{kj} \right)^2}{R_{kj}^2}-1 \right)\left(\frac{x_k}{R_{kj}}+\frac{\dot{x}_k}{c}\right)
-\frac{\left(\Delta Y_{kj}\right)^2}{R_{kj} c^2}\ddot{x}_k
+3\frac{\Delta X_{kj} \Delta Y_{kj}}{R_{kj}^3}y_k
+3\frac{\Delta X_{kj} \Delta Y_{kj}}{R_{kj}^2 c}\dot{y}_k
+\frac{\Delta X_{kj} \Delta Y_{kj}}{R_{kj} c^2}\ddot{y}_k \right],\\
&E_{kjy}\cong \frac{-N_k e}{4\pi\varepsilon_0 R_{kj}^2}
\left[\left( \frac{3\left(\Delta Y_{kj} \right)^2}{R_{kj}^2}-1 \right)\left(\frac{y_k}{R_{kj}}+\frac{\dot{y}_k}{c}\right)
-\frac{\left(\Delta X_{kj}\right)^2}{R_{kj} c^2}\ddot{y}_k
+3\frac{\Delta X_{kj} \Delta Y_{kj}}{R_{kj}^3}x_k
+3\frac{\Delta X_{kj} \Delta Y_{kj}}{R_{kj}^2 c}\dot{x}_k
+\frac{\Delta X_{kj} \Delta Y_{kj}}{R_{kj} c^2}\ddot{x}_k \right],\\
&j=1,~2,~...,~n.
\end{aligned}
\label{eq:Exy}
\end{equation}
Here, $\Delta X_{kj}=X_k-X_j, \Delta Y_{kj}=Y_k-Y_j$, and the higher order terms are ignored due to the assumption that the oscillation is of small amplitude and that no relativistic effects are considered, i.e., $x_k/R_{kj}\ll 1$ and $v/c\ll 1$.
$N_k$ is the free electron number of the $k$th MNP. The electrons in an MNP oscillate collectively due to the surface plasmon effect. Therefore, we use $N_k$ to represent the contributions of all the oscillating electrons of the $k$th MNP.

Substitute Eq. \ref{eq:Exy} into Eq. \ref{eq:xy0}, we obtain the oscillators' equations of the many-body system:
\begin{equation}
\centering
\begin{aligned}
&\ddot{x}_j+\beta_{0xj}\dot{x}_j+\omega_{0xj}^2 x_j
+\sum_{k\neq j}^{n}\left(
g_{kjxx}^2 x_k+\gamma_{kjxx}\dot{x}_k+\eta_{kjxx}\ddot{x}_k
+g_{kjyx}^2 y_k+\gamma_{kjyx}\dot{y}_k+\eta_{kjyx}\ddot{y}_k
\right)=C_{jx}\mathrm{exp}(-\mathrm{i}\omega_{ex}t),\\
&\ddot{y}_j+\beta_{0yj}\dot{y}_j+\omega_{0yj}^2 y_j
+\sum_{k\neq j}^{n}\left(
g_{kjyy}^2 y_k+\gamma_{kjyy}\dot{y}_k+\eta_{kjyy}\ddot{y}_k
+g_{kjxy}^2 x_k+\gamma_{kjxy}\dot{x}_k+\eta_{kjxy}\ddot{x}_k
\right)=C_{jy}\mathrm{exp}(-\mathrm{i}\omega_{ex}t),\\
&j=1,~2,~...,~n.
\end{aligned}
\label{eq:xy2}
\end{equation}
The coupling coefficients tensors can be written as:
\begin{equation}
\centering
\begin{aligned}
&G_{kj}=
\left(
\begin{matrix}
g_{kjxx}^2 & g_{kjxy}^2 \\
g_{kjyx}^2 & g_{kjyy}^2
\end{matrix}
\right)
=-N_k \alpha \frac{\hbar \omega_{kj}}{m_e c^2}\omega_{kj}^2
\left(
\begin{matrix}
3\cos^2\theta_{kj}-1 & 3\cos\theta_{kj} \sin\theta_{kj} \\
3\cos\theta_{kj} \sin\theta_{kj} & 3\sin^2\theta_{kj}-1
\end{matrix}
\right),\\
&\gamma_{kj}=
\left(
\begin{matrix}
\gamma_{kjxx} & \gamma_{kjxy} \\
\gamma_{kjyx} & \gamma_{kjyy}
\end{matrix}
\right)
=-N_k \alpha \frac{\hbar \omega_{kj}}{m_e c^2}\omega_{kj}
\left(
\begin{matrix}
3\cos^2\theta_{kj}-1 & 3\cos\theta_{kj} \sin\theta_{kj} \\
3\cos\theta_{kj} \sin\theta_{kj} & 3\sin^2\theta_{kj}-1
\end{matrix}
\right),\\
&\eta_{kj}=
\left(
\begin{matrix}
\eta_{kjxx} & \eta_{kjxy} \\
\eta_{kjyx} & \eta_{kjyy}
\end{matrix}
\right)
=-N_k \alpha \frac{\hbar \omega_{kj}}{m_e c^2}
\left(
\begin{matrix}
-\sin^2\theta_{kj} & \cos\theta_{kj} \sin\theta_{kj} \\
\cos\theta_{kj} \sin\theta_{kj} & -\cos^2\theta_{kj}
\end{matrix}
\right),\\
&j,~k=1,~2,~...,~n,~\mathrm{and}~j\neq k.
\end{aligned}
\label{eq:tensor1}
\end{equation}
Here, $\alpha=\frac{e^2}{4\pi \varepsilon_0 c \hbar}$ is the fine--structure constant, $\hbar$ is the reduced Planck constant, $\omega_{kj}=c/R_{kj}$, $\cos \theta_{kj}=\Delta X_{kj}/R_{kj}$, and $\sin \theta_{kj}=\Delta Y_{kj}/R_{kj}$.
From Eq. \ref{eq:xy2} we find that the oscillations in both $x$ and $y$ directions of the $j$th oscillator are affected by the oscillations in both $x$ and $y$ directions of the $k$th oscillator ($j\neq k$), which makes the problem more complicated.
In order to solve the equations, assume $x_j=A_{2j-1}\mathrm{exp}(-\mathrm{i}\omega_{ex}t)$ and $y_j=A_{2j}\mathrm{exp}(-\mathrm{i}\omega_{ex}t)$; define $K_{2j-1}=C_{jx}$ and $K_{2j}=C_{jy}$ to reorder $\left\{C_{jx},C_{jy}\right\}$; substitute them into Eq. \ref{eq:xy2}. As a result, we obtain the equations in matrix form:
\begin{equation}
\begin{aligned}
&\left(
\begin{matrix}
D_{11} & D_{12} & ... & D_{1N}\\
D_{21} & D_{22} & ... & D_{2N}\\
...\\
D_{N1} & D_{N2} & ... & D_{NN}\\
\end{matrix}
\right)
\left(
\begin{matrix}
A_{1}\\
A_{2}\\
...\\
A_{N}
\end{matrix}
\right)
=\left(
\begin{matrix}
K_{1}\\
K_{2}\\
...\\
K_{N}
\end{matrix}
\right),~~\mathrm{with}~N=2n,\\
&\mathrm{or~in~simple~form:}~DA=K.
\end{aligned}
\label{eq:matrix}
\end{equation}
Here, the elements of matrix $D$ can be expressed as:
\begin{equation}
\begin{aligned}
&D_{j_1 j_2}=\omega_{0xj}^2-\mathrm{i}\beta_{0xj}\omega_{ex}-\omega_{ex}^2, ~\mathrm{case}~ j_1=j_2=2j-1,\\
&D_{j_1 j_2}=\omega_{0yj}^2-\mathrm{i}\beta_{0yj}\omega_{ex}-\omega_{ex}^2,~\mathrm{case}~j_1=j_2=2j,\\
&D_{j_1 j_2}=g_{kjxx}^2-\mathrm{i}\gamma_{kjxx}\omega_{ex}-\eta_{kjxx}\omega_{ex}^2, ~\mathrm{case}~ j_1=2j-1,~ j_2=2k-1,~ j_1\neq j_2,\\
&D_{j_1 j_2}=g_{kjyx}^2-\mathrm{i}\gamma_{kjyx}\omega_{ex}-\eta_{kjyx}\omega_{ex}^2,~\mathrm{case}~j_1=2j-1,~ j_2=2k, \\
&D_{j_1 j_2}=g_{kjyy}^2-\mathrm{i}\gamma_{kjyy}\omega_{ex}-\eta_{kjyy}\omega_{ex}^2, ~\mathrm{case}~j_1=2j,~ j_2=2k,~ j_1\neq j_2,\\
&D_{j_1 j_2}=g_{kjxy}^2-\mathrm{i}\gamma_{kjxy}\omega_{ex}-\eta_{kjxy}\omega_{ex}^2,~\mathrm{case}~j_1=2j,~ j_2=2k-1,\\
\end{aligned}
\label{eq:D}
\end{equation}
Therefore, the solutions of $A_{j}$ are:
\begin{equation}
A=D^{-1}K,
\end{equation}
where $D^{-1}$ is the the inverse of matrix $D$. The far field radiation is considered here. We calculate the electric field at the position $\textbf{d}=\left(X_d,Y_d,Z_d\right)$ introduced by the coupled oscillators which are excited by the external light. According to Eq. \ref{eq:Ert}, the total far field radiation is derived as:
\begin{equation}
\begin{aligned}
&\textbf{E}^{(tot)}=-\frac{e}{4 \pi \varepsilon_0}\sum_{j=1}^{n} N_j \frac{r_j}{\left(\textbf{r}_j \cdot \textbf{u}_j \right)^3} \textbf{r}_j \times \left(\textbf{u}_j\times \textbf{a}_j \right),\\
&\textbf{B}^{(tot)}=-\frac{e}{4 \pi \varepsilon_0 c}\sum_{j=1}^{n} N_j \frac{1}{\left(\textbf{r}_j \cdot \textbf{u}_j \right)^3} \textbf{r}_j \times \left[ \textbf{r}_j \times \left(\textbf{u}_j\times \textbf{a}_j \right)\right],\\
&\textbf{S}\equiv \frac{1}{\mu_0}\left[\textbf{E}^{(tot)}\times \textbf{B}^{(tot)}\right].
\end{aligned}
\label{eq:Efar}
\end{equation}
Here, $\textbf{E}^{(tot)}$ and $\textbf{B}^{(tot)}$ represent the total electric field and the total magnetic field at the position $\textbf{d}$ introduced by the system, while $\textbf{S}$ is the Poynting vector; $\textbf{r}_j=\textbf{d}-\textbf{R}_j=\left(X_d-X_j,Y_d-Y_j,Z_d\right)$, $\textbf{u}_j=c\textbf{r}_j/r_j-\textbf{v}_j$; $r_j=\left|\textbf{r}_j\right|$,  $\textbf{v}_j=\left(\dot{x}_j,\dot{y}_j\right)$, $\textbf{a}_j=\left(\ddot{x}_j,\ddot{y}_j\right)$.

In a simple case, we assume that $d=\left|\textbf{d}\right|\gg \mathrm{max}\left(R_{kj}\right)$, where $\mathrm{max}\left(R_{kj}\right)$  is the maximum of $R_{kj}$ for arbitrary $j\neq k$, and the vector from the system to the field point is perpendicular to $xy$ plane, i.e., $\left| \textbf{r}_j\right|\cong \left| Z_d\right| $ for arbitrary $j$. As a result, the expressions of the electric field in Eq. \ref{eq:Efar} can be approximately simplified as:
\begin{equation}
\begin{aligned}
E_x^{(tot)}&\cong \sum_{j=1}^{n} \frac{N_j e}{4 \pi \varepsilon_0 c^2 r} \ddot{x}_j=\frac{- e \omega_{ex}^2}{4 \pi \varepsilon_0 c^2 r}\left(\sum_{j=1}^{n} N_j A_{2j-1}\right)\mathrm{exp}(-\mathrm{i}\omega_{ex}t)
:=A_x(\omega_{ex}) \mathrm{exp}(-\mathrm{i}\omega_{ex}t),\\
E_y^{(tot)}&\cong \sum_{j=1}^{n} \frac{N_j e}{4 \pi \varepsilon_0 c^2 r} \ddot{y}_j=\frac{- e \omega_{ex}^2}{4 \pi \varepsilon_0 c^2 r}\left(\sum_{j=1}^{n} N_j A_{2j}\right)\mathrm{exp}(-\mathrm{i}\omega_{ex}t)
:=A_y(\omega_{ex}) \mathrm{exp}(-\mathrm{i}\omega_{ex}t),\\
E_z^{(tot)}&\cong 0.
\end{aligned}
\label{eq:EfarA}
\end{equation}
Here, $r=\left|Z_d\right|$ is defined as the distance between the system and the field point in this simple case.
Therefore the scattering spectrum of the system can be evaluated by:
\begin{equation}
I_{sca}(\omega)= \left|A_x\left(\omega_{ex}=\omega\right)\right|^2
+\left|A_y\left(\omega_{ex}=\omega\right)\right|^2.
\end{equation}

\section{\label{sec:Results}Results and Discussions}
\begin{figure}[h]
\includegraphics[width=0.48\textwidth]{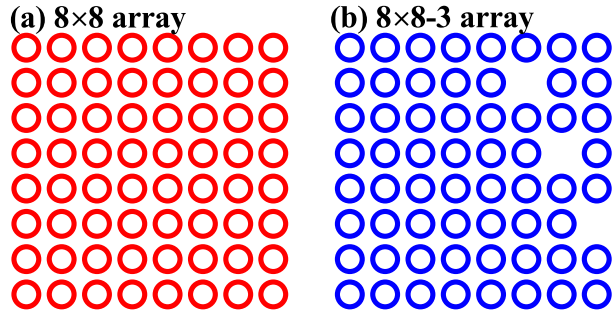}
\caption{\label{fig:Conf8} Schematic of the array which is in $xy$ plane. It contains $n=8\times 8=64$ MNPs with the same parameters, i.e., the free electron number is $10^5$, the resonance wavelength is 550 nm, and the linewidth is 80 nm. (a) represents the entire array, while (b) represents the defective array in the absence of three MNPs.
}
\end{figure}

\begin{figure}[h]
\includegraphics[width=0.48\textwidth]{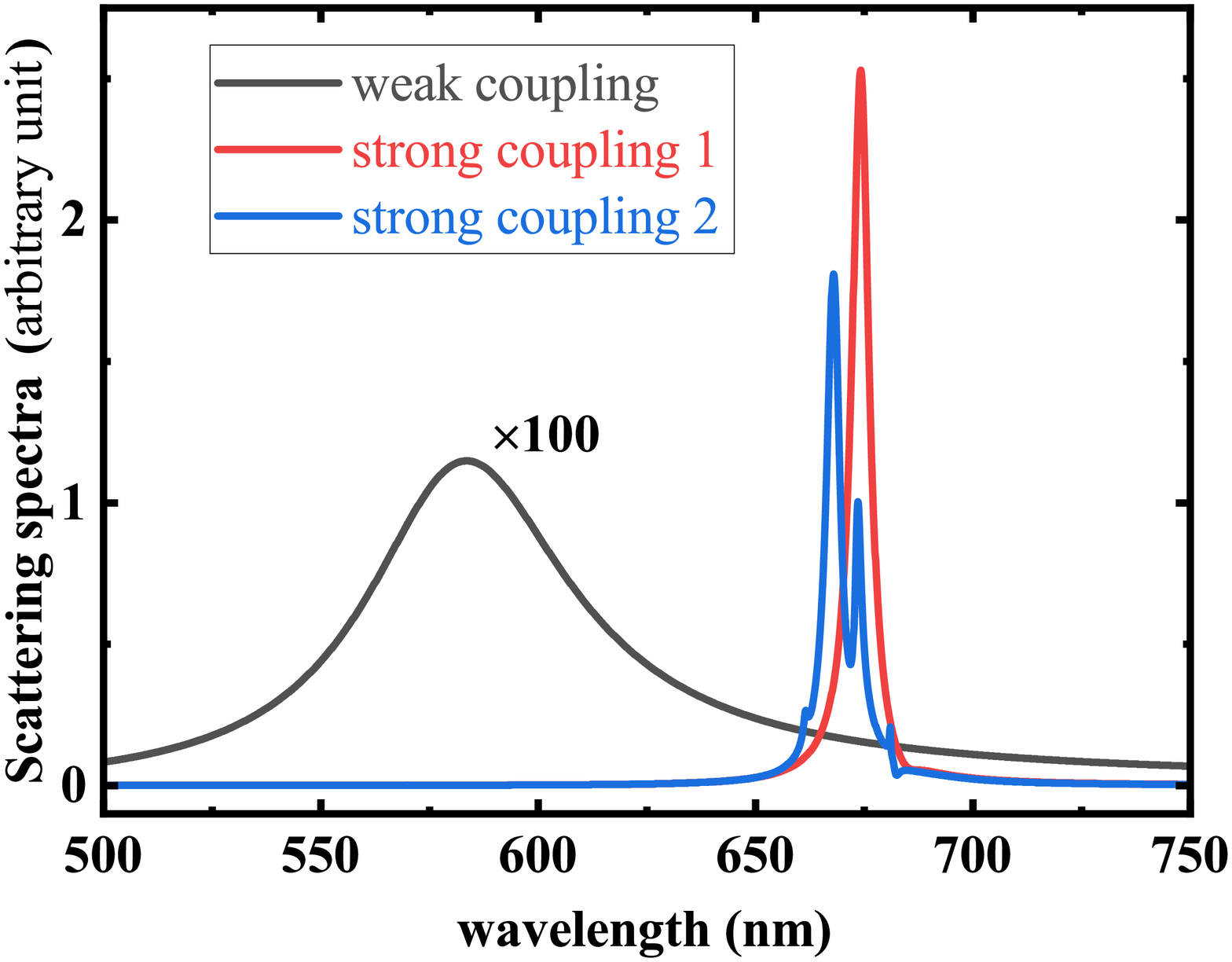}
\caption{\label{fig:Isca} Scattering spectra of the array. Black and red curves represent the entire arrays of weak and strong coupling, respectively. Red curve represents the defective array of strong coupling. The missing three MNPs are shown in Fig. \ref{fig:Conf8}. The values of the black curve are magnified by a factor of 100.
}
\end{figure}

Here, we show an example of an $8\times 8$ array with the period $p$, in which each MNP has the same optical property, as shown in Fig. \ref{fig:Conf8}, where the red ones represent the entire array while the blue ones represent the defective array.
Fig. \ref{fig:Isca} shows the scattering spectra of the array in different cases. It is obvious that the scattering intensity of weakly coupled array is much less than the one of strongly coupled array, while the linewidth of the former is much larger than the one of the latter. Here, weak coupling corresponds to $p=50$ nm and strong coupling corresponds to $p=30$ nm. Furthermore, the defective array shows multiple peaks rather than the entire array that shows only one peak. It indicates that the defective array supports more electromagnetic modes, which is beneficial for controlling light at subwavelength scale.

\begin{figure}[h]
\includegraphics[width=0.48\textwidth]{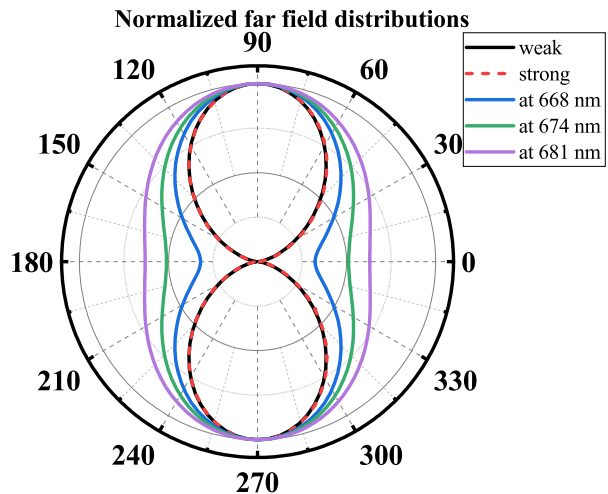}
\caption{\label{fig:FarField} Normalized $xz$ plane far field radiation distributions of different cases with $x$ polarized external light. Black and red curves represent weak and strong coupling, which correspond to $p=50$ nm and $p=30$ nm, excited at their resonance wavelengths, respectively. Blue, green, and purple curves represent strong coupling excited at 668 nm, 674 nm, and 681 nm, respectively.
}
\end{figure}

Fig. \ref{fig:FarField} shows the normalized $xz$ plane far field radiation distributions of the arrays. Obviously, the entire arrays in both weak and strong coupling cases illustrate an ``8--shaped'' distributions, which is like the far field distributions of a dipole oscillating along $x$ axis. However, when it comes to the defective array, the distributions are different, i.e., they are tuned varying with the excitation wavelength. The far field radiation intensity in $x$ increases as the excitation wavelength increases. This phenomenon illustrates that the defective array can control far field radiation distributions by varying the excitation wavelength. 

From the above example, we expect that other configurations of the MNPs can also control the light emissions. Therefore, in order to obtain the optical properties that we want, careful design of the configuration should be made. We look forward to more novel configurations which present novel optical properties.


\section*{\label{sec:Conclusion}Conclusions}
In conclusion, we develop a classic model to reveal the optical properties of arbitrary configurations of MNPs in plane, considering the variation of the external light. Scattering spectra and far field radiation distributions of an example are shown. It is convenient to evaluate the optical properties of coupled MNPs using this model. We hope this work is helpful to the applications in nanophotonics.

\section*{\label{sec:Acknowldegment}Acknowledgment}
This work was supported by the Fundamental Research Funds for the Central Universities (Grant No. FRF-TP-20-075A1).

\section*{Disclosures}
The author declares no conflicts of interest.

\section*{Data availability}
The data that support the findings of this study are available from the corresponding author upon reasonable request.

\section*{\label{sec:Ref}References}

\bibliography{TBW_Cheng}

\providecommand{\latin}[1]{#1}
\makeatletter
\providecommand{\doi}
  {\begingroup\let\do\@makeother\dospecials
  \catcode`\{=1 \catcode`\}=2 \doi@aux}
\providecommand{\doi@aux}[1]{\endgroup\texttt{#1}}
\makeatother
\providecommand*\mcitethebibliography{\thebibliography}
\csname @ifundefined\endcsname{endmcitethebibliography}
  {\let\endmcitethebibliography\endthebibliography}{}
\begin{mcitethebibliography}{8}
\providecommand*\natexlab[1]{#1}
\providecommand*\mciteSetBstSublistMode[1]{}
\providecommand*\mciteSetBstMaxWidthForm[2]{}
\providecommand*\mciteBstWouldAddEndPuncttrue
  {\def\EndOfBibitem{\unskip.}}
\providecommand*\mciteBstWouldAddEndPunctfalse
  {\let\EndOfBibitem\relax}
\providecommand*\mciteSetBstMidEndSepPunct[3]{}
\providecommand*\mciteSetBstSublistLabelBeginEnd[3]{}
\providecommand*\EndOfBibitem{}
\mciteSetBstSublistMode{f}
\mciteSetBstMaxWidthForm{subitem}{(\alph{mcitesubitemcount})}
\mciteSetBstSublistLabelBeginEnd
  {\mcitemaxwidthsubitemform\space}
  {\relax}
  {\relax}

\bibitem[Garrido~Alzar \latin{et~al.}(2002)Garrido~Alzar, Martinez, and
  Nussenzveig]{couple2}
Garrido~Alzar,~C.~L.; Martinez,~M. A.~G.; Nussenzveig,~P. Classical analog of
  electromagnetically induced transparency. \emph{American Journal of Physics}
  \textbf{2002}, \emph{70}, 37--41\relax
\mciteBstWouldAddEndPuncttrue
\mciteSetBstMidEndSepPunct{\mcitedefaultmidpunct}
{\mcitedefaultendpunct}{\mcitedefaultseppunct}\relax
\EndOfBibitem
\bibitem[Prodan \latin{et~al.}(2003)Prodan, Radloff, Halas, and
  Nordlander]{couple3}
Prodan,~E.; Radloff,~C.; Halas,~N.~J.; Nordlander,~P. A Hybridization Model for
  the Plasmon Response of Complex Nanostructures. \emph{Science} \textbf{2003},
  \emph{302}, 419--422\relax
\mciteBstWouldAddEndPuncttrue
\mciteSetBstMidEndSepPunct{\mcitedefaultmidpunct}
{\mcitedefaultendpunct}{\mcitedefaultseppunct}\relax
\EndOfBibitem
\bibitem[Davis \latin{et~al.}(2010)Davis, Gómez, and Vernon]{couple4}
Davis,~T.; Gómez,~D.; Vernon,~K. Simple model for the hybridization of surface
  plasmon resonances in metallic nanoparticles. \emph{Nano letters}
  \textbf{2010}, \emph{10}, 2618—2625\relax
\mciteBstWouldAddEndPuncttrue
\mciteSetBstMidEndSepPunct{\mcitedefaultmidpunct}
{\mcitedefaultendpunct}{\mcitedefaultseppunct}\relax
\EndOfBibitem
\bibitem[G\'omez \latin{et~al.}(2012)G\'omez, Roberts, Davis, and
  Vernon]{couple5}
G\'omez,~D.~E.; Roberts,~A.; Davis,~T.~J.; Vernon,~K.~C. Surface plasmon
  hybridization and exciton coupling. \emph{Phys. Rev. B} \textbf{2012},
  \emph{86}, 035411\relax
\mciteBstWouldAddEndPuncttrue
\mciteSetBstMidEndSepPunct{\mcitedefaultmidpunct}
{\mcitedefaultendpunct}{\mcitedefaultseppunct}\relax
\EndOfBibitem
\bibitem[Downing and Weick(2020)Downing, and Weick]{couple6}
Downing,~C.~A.; Weick,~G. Plasmonic modes in cylindrical nanoparticles and
  dimers. \emph{Proceedings of the Royal Society A: Mathematical, Physical and
  Engineering Sciences} \textbf{2020}, \emph{476}, 20200530\relax
\mciteBstWouldAddEndPuncttrue
\mciteSetBstMidEndSepPunct{\mcitedefaultmidpunct}
{\mcitedefaultendpunct}{\mcitedefaultseppunct}\relax
\EndOfBibitem
\bibitem[Cheng and Sun(2022)Cheng, and Sun]{couple0}
Cheng,~Y.; Sun,~M. Unified Treatment for Photoluminescence and Scattering of
  Coupled Metallic Nanostructures: I. Two-Body System. \emph{New J. Phys.}
  \textbf{2022}, \emph{24}, 033026\relax
\mciteBstWouldAddEndPuncttrue
\mciteSetBstMidEndSepPunct{\mcitedefaultmidpunct}
{\mcitedefaultendpunct}{\mcitedefaultseppunct}\relax
\EndOfBibitem
\bibitem[Cheng and Sun(2022)Cheng, and Sun]{couple1}
Cheng,~Y.; Sun,~M. Unified Treatment for Scattering, Absorption, and
  Photoluminescence of coupled Metallic Nanoparticles with Vertical Polarized
  Excitation. \emph{arXiv:2211.04469 [physics.optics]} \textbf{2022}, \relax
\mciteBstWouldAddEndPunctfalse
\mciteSetBstMidEndSepPunct{\mcitedefaultmidpunct}
{}{\mcitedefaultseppunct}\relax
\EndOfBibitem
\bibitem[Griffiths(2013)]{Griffiths}
Griffiths,~D.~J. \emph{Introduction to Electrodynamics (4rd Edition)}; Pearson:
  Cambridge, U.K., 2013\relax
\mciteBstWouldAddEndPuncttrue
\mciteSetBstMidEndSepPunct{\mcitedefaultmidpunct}
{\mcitedefaultendpunct}{\mcitedefaultseppunct}\relax
\EndOfBibitem
\end{mcitethebibliography}
\bibliographystyle{achemso}

\end{document}